\documentclass[cup6b]{cupbook}
\usepackage{graphicx}
\usepackage{natbib}
\title[Rome, Italy, 27--30 April 2009]
      {The coming of age of X-ray polarimetry}
\author{}
\date{}
\begin{document}
\pagenumbering{arabic}


\author[Stefano Covino]{Stefano Covino \\  INAF / Brera Astronomical Observatory, Via Bianchi 46, 23807, Merate (LC) - Italy}
\chapter{GRB Afterglow Polarimetry\\Past, Present and Future}

\abstract{Gamma-ray bursts and their afterglows are thought to be produced by an ultrarelativistic jet. One of the most 
important open questions is the outflow composition: the energy may be carried out from the central source either as 
kinetic energy (of baryons and/or pairs), or in electromagnetic form (Poynting flux). While the total observable flux 
may be indistinguishable in both cases, its polarization properties are expected to differ markedly. The later time 
evolution of afterglow polarization is also a powerful diagnostic of the jet geometry. Again, with subtle and hardly
detectable differences in the output flux, we have distinct polarization predictions.}

\section{Introduction}

Polarimetry is a powerful diagnostic tool to study spatially unresolved sources at cosmological distances, such as gamma-ray burst
(GRB) afterglows. Radiation mechanisms that produce similar spectra can be disentangled by means of their polarization signatures. Also,
polarization provides unique insights into the geometry of the source, which remains hidden in the integrated light.

\begin{figure}
\begin{tabular}{cc}
\includegraphics[scale=0.65]{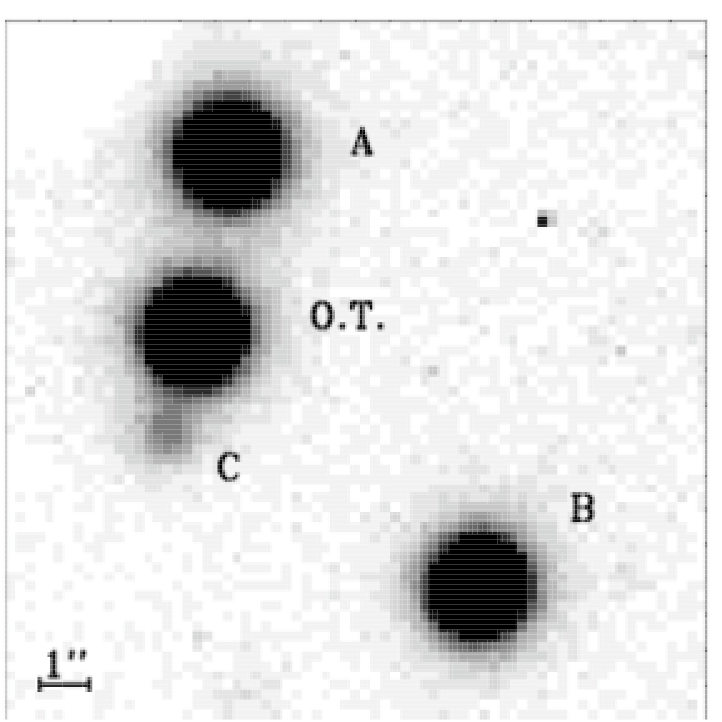} &
\includegraphics[scale=0.35]{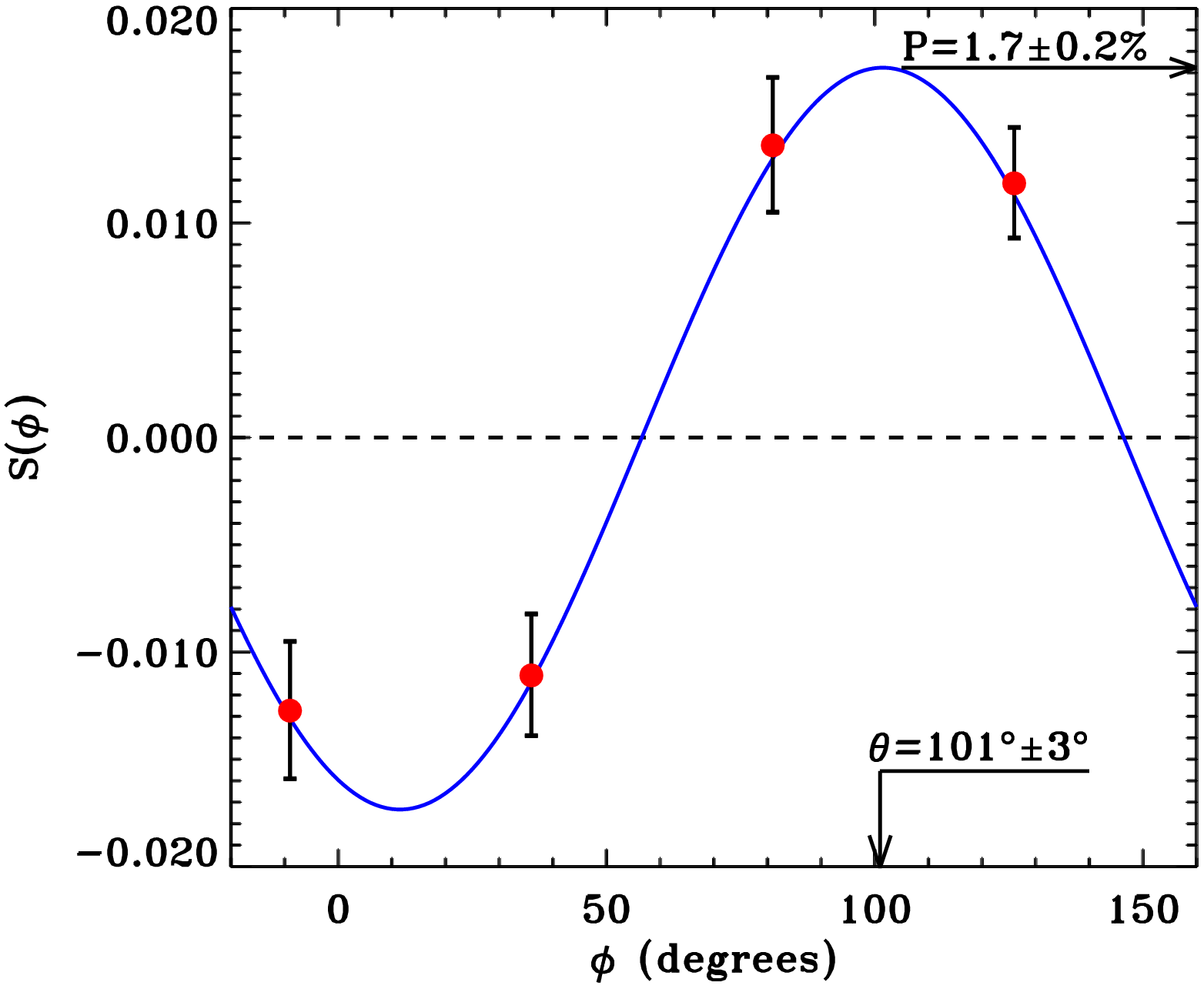} \\
\end{tabular}
\caption{The field of GRB\,990510 observed by the ESO-VLT equipped with FORS1 in the R band (left). The net polarization of GRB\,990510 (right). From \cite{Co99}.}
\label{fig:pol}
\end{figure}

Historically, essentially all interpretative studies about GRB afterglow polarimetry have been based on the cosmological fireball model \cite{Pir99,ZhaMe03}, which we will also use as a reference for our discussion. Afterglow polarization studies have indeed the advantage 
that different models are often almost indistinguishable in term of radiation output in the optical, but produce markedly distinct 
predictions about polarization. 

In this proceeding, we will briefly review what we have derived by optical afterglow polarimetric observations in Sect.\,\ref{sec:past} 
and discuss the most recent development in the field in Sect.\,\ref{sec:early}. For a deeper discussion about the physical ingredients
generating a polarized flux in GRB afterglow radiation one can refer to other proceedings in this volume \cite{Lau09,Laz09,Fa09}.

\section{What have we learnt so far?}
\label{sec:past}

We report below what we consider the three most important achievements obtained by 
afterglow polarimetric observation in GRB research. Generally speaking, two general families 
of models have been developed to explain  why GRB afterglows can be polarized and the 
time evolution of polarization. One possibility  is that the emission originates in causally disconnected 
regions of highly ordered magnetic field, each producing polarization almost at the maximum 
degree. \cite{GW99} predicted a $\sim 10\%$ polarization. If the regions have a
statistical distribution of energies, the position angle can be different at
various wavelengths. This value is greater than that observed in many GRB afterglows \cite{Co04}
as most of the positive detections so far derived are below $\sim 3$\%. In an alternative scenario first introduced by \cite{Melo99} and then developed by \cite{GhLa99,Sa99} the magnetic field is ordered in the plane of the shock. In a spherical
fireball, such a field configuration would give null polarization, but if a
collimated fireball is observed off-axis (as it is most probable), a small
degree of polarization would be predicted, with a well defined temporal
evolution. Here the ultrarelativistic motion toward the observer and the physical beaming of the
outflow are fundamental ingredients.

\subsection{GRB afterglows polarization}

After a few unfruitful attempts which \cite{Hj99}, and not by chance as soon as the first unit of the ESO-VLT become operational, a low
although highly significant polarization for the afterglow of GRB\,990510 (Fig.\,\ref{fig:pol}) was successfully detected 
for the first time \cite{Co99,Wi99}. This simple observational finding carried already a lot of information. First of all, 
the detection of polarized flux from a GRB afterglow can and has been considered a clear signature for synchrotron 
emission, although various alternative explanations indeed exist. In general, the
detected polarization ($1.7\% \pm 0.2\%$) would require emission processes
involving particle acceleration. In the external shock phenomenon we have particle acceleration at the shock front
and once we consider the ultrarelativistic motion toward the observer and the physical beaming of the outflow
some level of polarization in the afterglows is naturally predicted. It is possible to have some degree
of polarization adopting other scenarios, however in no case a polarized flux is a natural output of the model, as it is 
for the cosmological fireball model. To my knowledge, this is still one of the most convincing, although admittedly often unrecorded,
observational proof supporting the standard afterglow model.

\subsection{Afterglow polarization variable in time} 

\begin{figure}
\includegraphics[scale=0.5]{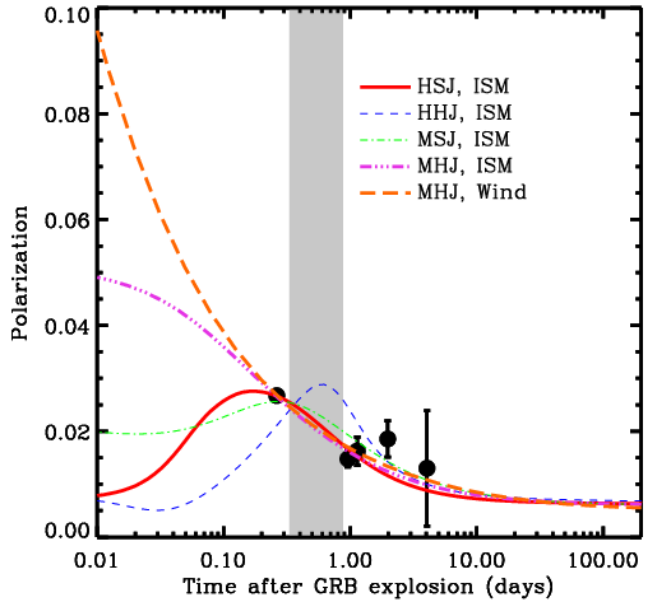}
\caption{Polarimetric curve of GRB\,080213. The observations are compared to predictions for several families of 
models as discussed in \cite{Laz04}. The shaded area shows where a break, possibly a jet break, was observed during
the optical afterglow evolution \cite{Co03}.}
\label{fig:curv}
\end{figure}

The detection of varying polarization on time scales comparable to those of the afterglow evolution, 
immediately implies that the observed polarization is intrinsic to the
source and not, for instance, due to scattering against material along the line of
sight. The first convincing evidence of time-varying polarization was
obtained for GRB\,020813 \cite{Ba03,Laz04}, where a decrease of the polarization degree
from $\sim 3\%$ down to less than $1\%$ (Fig.\,\ref{fig:curv}), with constant position angle,
was recorded from a few hours to half a day after the burst. Evolution was also singled out in GRB\,021004 \cite{Laz03}.

The most striking example is however GRB\,030329. Due to its relatively small distance, 
a very high quality dataset was obtained covering more than two weeks \cite{Gre03}. 
Strong, somewhat erratic, variations of the polarization degree and position angle during 
the afterglow evolution were singled out. Polarization variations occurred on a time scale
comparable to that of the afterglow flux variability, offering a direct link between the two phenomena \cite{GrKo03}
although in the case of GRB\,030329 the late-time rise of the supernova (associated to the GRB) component
played an important role.

\subsection{Afterglow polarization and the geometry of GRB jets}

Since GRB sources are unresolved, any model for producing polarization requires some 
kind of anisotropy in the emitting fluid. The simplest configuration envisages emission from 
homogenous jets observed off-axis. In this case, as shown by \cite{GhLa99,Sar99}, the polarization 
time evolution presents two maxima, reaching a zero level in
between, where a flip of the polarization angle by $90^\circ$ also
occurs. A clear prediction in principle easy to test with observations.
More complex jet structures have also been proposed, i.e. in which the energy content 
per solid angle is decreasing towards the wings of the jet. Such configurations may allow
for a unified view of GRBs, in which differences among events arise only
(or mostly) from the orientation of the observer with respect to the jet
core. The polarization behaviour is in this case markedly different,
with a single broad maximum \cite{Ros04,Laz04}. 

Up to now, a full set of (late time) 
observations of polarization evolution could effectively be compared with 
models only in the case of GRB\,020813. The main result of accurate modeling \cite{Laz04}
is that homogeneous jet model predictions with shock  generated magnetic fields are in 
clear disagreement with the observations.  This is one of the strongest direct observational evidences 
against the homogeneous jet scenario so far obtained (see also the case of
GRB\,030328, \cite{Mai06}).

\section{\textit{Swift} and the early afterglow}
\label{sec:early}

Time-resolved polarimetric observations of late-time (later than about 1\,hour) afterglow are extremely 
demanding, even for 8\,m class telescopes, due to the low polarization detected and the rapid fading 
of the afteeglows. Moreover, the complexity
of afterglow behaviors compared to the theoretical predictions made difficult to apply further observational
tests and derive unambiguous answers. 

However, after the launch of the \textit{Swift} satellite \cite{Geh04}, early afterglow observations become feasible
thanks also to the network of ground-based robotic telescopes devoted to GRB follow-up. Early afterglow observations 
can provide powerful diagnostics for many physical ingredients of GRB models, and again polarimetry can help to solve 
one of the hottest issues of GRB research.

Within the cosmological fireball model a hot fireball \cite{Pir99} expands driven 
by internal energy. An alternative scenario which attracted great theoretical interest has also 
been developed \cite{Uso92,Tho94,Lyu03,Lyu06,ZhKo05}, the ``electromagnetic outflow", where most of 
the energy is carried to large distances from the central source in electromagnetic
form (Poynting flux). Although dramatically different physics are involved, these two
scenarios may result in a similar radiation output. 

Things are different if polarization is considered. In the early phases optical emission can be 
generated by the forward shock, i.e. the afterglow, or by the reverse shock responsible 
for the optical flash. \cite{GrKo03} showed that the optical flash, if due to the reverse shock,
shares the same magnetic field configuration as the prompt emission, 
and therefore the same level of linear polarization. If the fireball is 
electromagnetically dominated, the first tens of minutes
of the afterglow may be $>$40\% polarized \cite{Laz06}. 
Optical flash polarization properties probe the magnetic field structure within the original 
outflow, while the afterglow emission probes the magnetic field structure in the shocked 
external medium, as well as the jet angular structure. Producing strong polarization in the optical flash 
requires a large scale ordered magnetic field possibly advected from the inner engine, while if the 
magnetic field is shock generated, no polarization is expected.

\subsection{Early afterglow observations}

\begin{figure}
\includegraphics[scale=0.75]{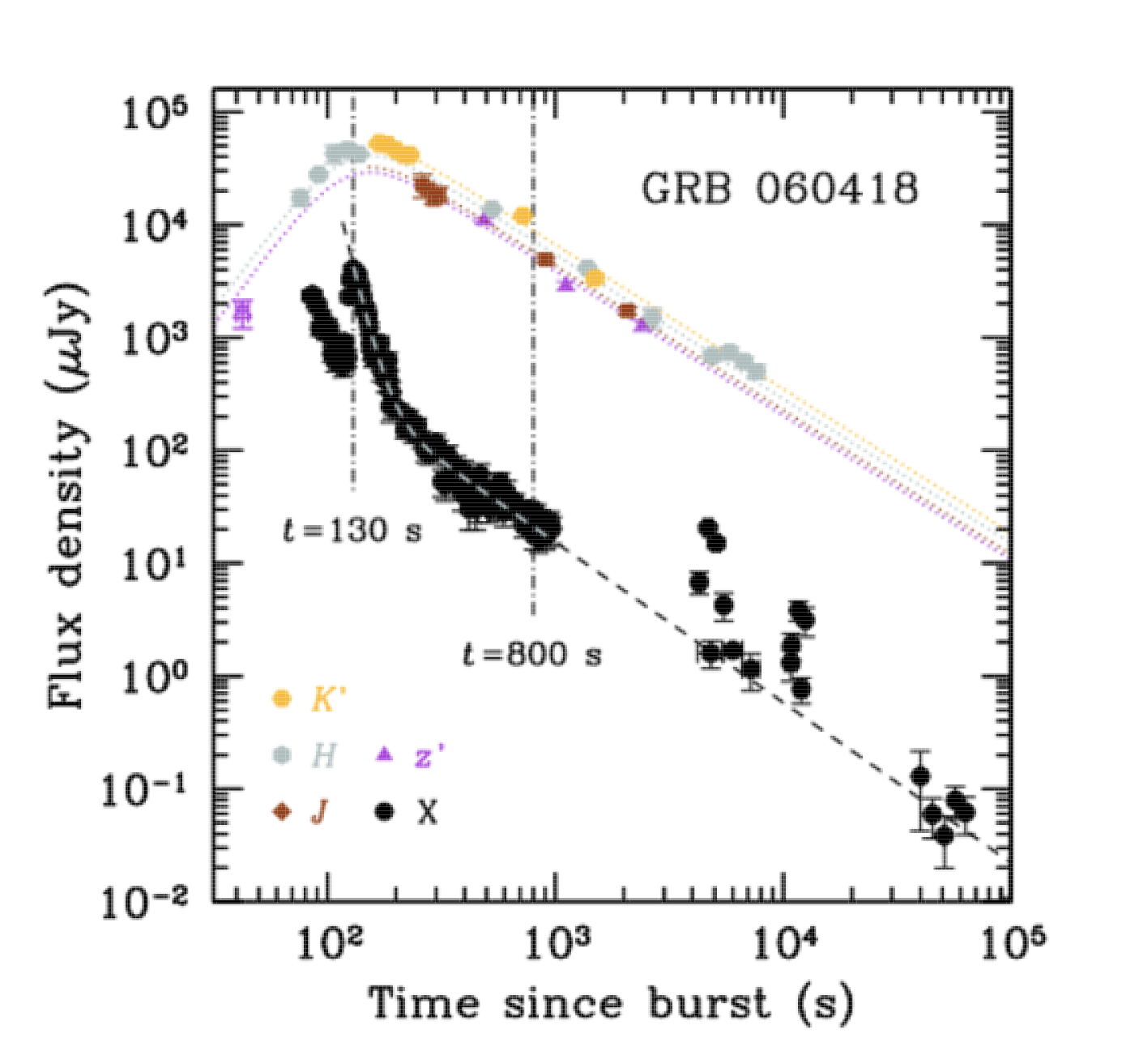}
\caption{Early afterglow near-infrared and X-ray light curves obtained by the REM telescope and by 
\textit{Swift} \cite{Mol07}. The Liverpool telescope polarimetric measurement \cite{Mun07} was carried out
about three minutes after the high-energy event, just after the peak of the near-infrared light curve which
was interpreted as the afterglow onset.}
\label{fig:earl}
\end{figure}

To date, the only early polarimetric measurement was performed by \cite{Mun07} deriving an 
8\% upper limit just after the onset of the afterglow of GRB\,060418 (Fig.\,\ref{fig:earl}). This result could 
strongly limit the possible role of magnetic fields in driving the outflow dynamics. However, in this case the 
optical emission is likely due to the forward shock only \cite{Mol07,JiFa07}, and the predicted polarization 
level depends on still poorly known details of the transfer of magnetic energy from the outflow to the 
shocked circumburst medium \cite{GrKo03,ZhKo05,FaPi06,Co07}, so that low or null polarization is still
compatible with the theoretical expectations and these measurements are not conclusive yet.

\begin{thereferences}{99}
\bibitem{Ba03} Barth et al. (2003) \textit{ApJ} \textbf{584}, L47 
\bibitem{Co99} Covino S. et al. (1999). \textit{A\&A} \textbf{348}, L1
\bibitem{Co03} Covino S. et al. (2003) \textbf{A\&A} \textit{404}, 5
\bibitem{Co04} Covino S. et al. (2004). \textit{ASPC} \textbf{312}, 169  
\bibitem{Co07} Covino S. (2007) \textit{Science} \textbf{315}, 1798
\bibitem{Fa09} Fan Y.Z. (2009) \textbf{this volume}
\bibitem{FaPi06} Fan Y.Z., Piran T. (2006) \textit{MNRAS} \textbf{369}, 197
\bibitem{JiFa07} Jin Z.P., Fan Y.Z. (2007) \textit{MNRAS} \textbf{378}, 1043
\bibitem{Geh04} Gehrels, N., et al. (2004). \textit{ApJ} \textbf{611} 1005
\bibitem{GhLa99} Ghisellini G., Lazzati D. (1999) \textit{MNRAS} \textbf{309} L7 
\bibitem{GrKo03} Granot J., K\"onigl A. (2003) \textit{ApJ} \textbf{594} L83
\bibitem{Gre03} Greiner et al. (2003) \textit{Nature} \textbf{426}, 157
\bibitem{GW99} Gruzinov A., Waxman E. (1999) \textit{ApJ} \textbf{511} 852
\bibitem{Hj99} Hjorth J., et al. (1999) \textit{Science} \textbf{283} 2073
\bibitem{Lau09} Laurent P. (2009) \textbf{this volume}
\bibitem{Laz06} Lazzati D. (2006) \textit{New Journ. Phys} \textbf{8}, 131
\bibitem{Laz09} Lazzati D. (2009) \textit{this volume} 
\bibitem{Laz03} Lazzati et al. (2003) \textit{A\&A} \textbf{410}, 823
\bibitem{Laz04} Lazzati et al. (2004) \textit{A\&A} \textbf{422}, 121
\bibitem{Lyu06} Lyutikov M. (2006) \textit{New Journ. Phys.} \textbf{8} 119
\bibitem{Lyu03} Lyutikov M., Pariev V.I., Blandford R.D. (2003) \textit{ApJ} \textbf{597}, 998
\bibitem{Mai06} Maiorano E. et al. (2006) \textit{A\&A} \textbf{455} 423
\bibitem{MeLo99} Medvedev M.V., Loeb A. (1999) \textit{ApJ} \textbf{526} 697
\bibitem{Mol07} Molinari E. et al. (2007) \textit{A\&A} \textbf{469}, 13
\bibitem{Mun07} Mundell C. et al. (2007) \textit{Science} \textbf{315}, 1822
\bibitem{Pir99} Piran T. (1999). \textit{Phys. Rep.} \textbf{314}, 575
\bibitem{Ros04} Rossi E. et al. (2004) \textit{MNRAS} \textbf{354} 86
\bibitem{Sar99} Sari R. (1999) \textit{ApJ} \textbf{524} L43
\bibitem{Tho94} Thompson C. (1994) \textit{MNRAS} \textbf{270}, 480
\bibitem{Uso92} Usov V.V. (1992) \textit{Nature} \textbf{357}, 472
\bibitem{Wi99} Wijers R. et al. (1999). \textit{ApJ} \textbf{523}, L33
\bibitem{ZhKo05} Zhang B., Kobayashi S. (2005) \textit{ApJ} \textbf{628}, 315
\bibitem{ZhaMe03} Zhang B., M\'esz\'aros P. (2003) \textit{Int. Journ. Mod. Phys. A} \textbf{19}, 2385
\end{thereferences}
\end{document}